\documentclass{iau} 
\usepackage{graphicx}

\title[AGB and post-AGB objects in the outer Galaxy] 
{AGB and post-AGB objects\\ in the outer Galaxy}

\author[Szczerba et al.] 
{Ryszard Szczerba$^1$,
Bosco H. K. Yung$^1$,
Marta Sewi{\l}o$^{2,3}$, \\
Natasza Si\'odmiak$^1$
\and  Agata Karska$^4$}

\affiliation{$^1$Nicolaus Copernicus Astronomical Centre, PAS, ul. Rabia\'nska 8, 87-100 Toru\'n, Poland 
\\[\affilskip]
$^2$NASA Goddard Space Flight Center, 8800 Greenbelt Rd., Greenbelt, MD 20771, USA\\ [\affilskip]
$^3$Astronomical Observatory of the Jagiellonian University, ul. Orla 171, 30-244 Krak\'ow, Poland \\[\affilskip]
$^4$Centre for Astronomy, Nicolaus Copernicus University, Faculty of Physics, Astronomy and Informatics,
ul. Grudzi\c{a}dzka 5, 87-100 Torun, Poland
}

\pubyear{2016}
\volume{323}  
\jname{Planetary Nebulae: Multi-Wavelength Probes of Stellar and Galactic Evolution}
\editors{X.W. Liu, L. Stanghellini \& A. Karakas, eds.}
\begin{document}

\maketitle

\begin{abstract}
We present the results of our search for low- and intermediate mass evolved stars in the outer Galaxy using AllWISE catalogue photometry. We show that the [3.4]$-$[12] vs. [4.6]$-$[22] colour-colour diagram is most suitable for separating C-rich/O-rich AGB and post-AGB star candidates. We are able to select 2,510 AGB and 24,821 post-AGB star candidates. However, the latter are severely mixed with the known young stellar objects in this diagram. 
\end{abstract}

\firstsection 
\section{Introduction}

We are conducting a systematic study of star formation in the outer Galaxy to uncover the population of intermediate- and low-mass young stellar objects (YSOs) and investigate the impact of the environment on the star formation process. 
We use the data from the ``\emph{Spitzer} Mapping of the Outer Galaxy'' survey (SMOG; PI: Sean Carey) that covered $\sim$24 deg$^{2}$ region in the outer Galaxy: {\it l} = (102$^{\circ}$, 109$^{\circ}$), {\it b} = ($-0.2^{\circ}$, $3.2^{\circ}$) in the IRAC 3.6--8.0 $\mu$m and MIPS 24 $\mu$m bands.
This relatively unstudied region, referred by us as ``L105'', have different environments and star formation activities. 

However, in the outer Galaxy including L105 we expect contamination from evolved stars such as asymptotic giant branch (AGB) stars, post-AGB objects and planetary nebulae (PNe). \cite[Szczerba et al. (2016)]{Szczerba_etal2016} concentrated on identifying the location of the low- and intermediate-mass evolved stars (AGBs, post-AGBs, and PNe) on the colour-colour diagram (CCD) based on the 2MASS and \emph{Spitzer} photometry ($K_{s}-[8.0]$ vs. $K_{s}-[24]$). This diagram allows us to separate C-rich and O-rich AGB stars quite effectively \cite[(Matsuura et al. 2014)]{Matsuura_etal2016}. Nonetheless, the total number of the SMOG sources with good photometric data in all these bands is quite limited (15,311 as compared to almost 3 millions sources detected in L105 by \emph{Spitzer} at the shorter wavelengths).  Therefore, we have found counterparts of the \emph{Spitzer} sources in the
Wide-field Infrared Survey Explorer (WISE) satellite survey at 3.4, 4.6, 12 and 22 $\mu$m and considered all CCDs based on different combinations of these bands.

\section{Results}

\begin{figure}[]
\begin{minipage}{15pc}
\vspace{-0.1pc}
\includegraphics[width=13.5pc]{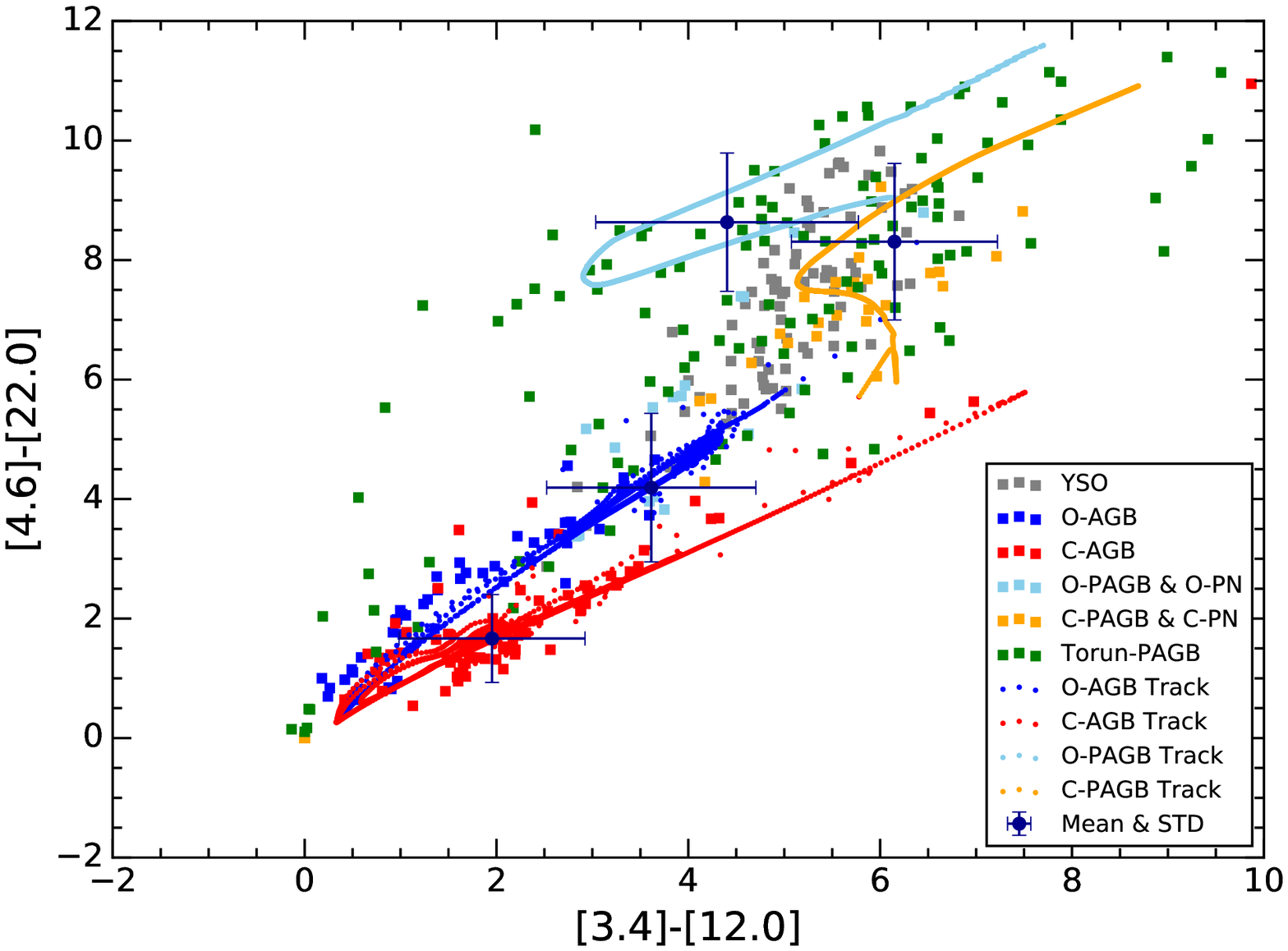}
\caption{\label{fig:hd} Comparison between HD models of AGB and post-AGB evolution with spectroscopically confirmed sources from the Magellanic Clouds and Galactic post-AGB objects. See the legend and text for details.}
\end{minipage} 
\hspace{1pc}
\begin{minipage}{15.5pc}
\includegraphics[width=16pc]{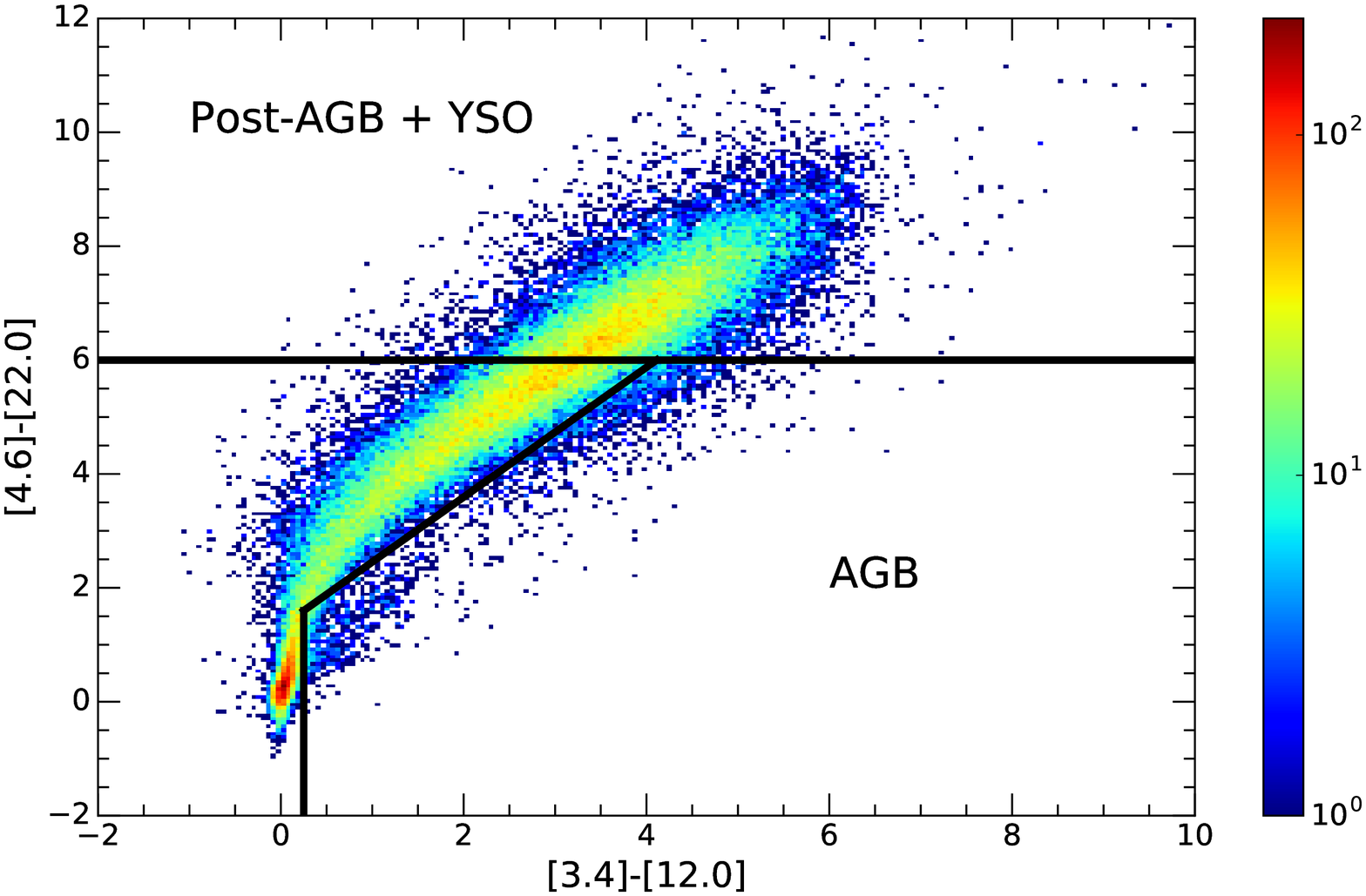}
\caption{\label{fig:hess} The Hess diagram for the SMOG sources. There are $200\times 200$ bins, and the colour 
indicates the number of sources 
in each bin. The black lines represent the AGB and post-AGB star selection criteria (see text).}
\end{minipage}
\end{figure}

Amongst a total of 15 CCDs, the best option for separating C-rich AGB and post-AGB stars from the O-rich ones is the $[3.4]-[12]$ vs. $[4.6]-[22]$ CCD. In Figure~\ref{fig:hd}, we show on this CCD the hydrodynamical (HD) models for gaseous dusty circumstellar shells around C-rich and O-rich stars during their final stages of AGB (red and blue dots, respectively) and post-AGB (orange and light blue dots, respectively) evolution \cite[(Steffen et al. 1998)]{Steffen_etal1998}. The mean and standard deviation (STD) of the model distributions are also shown by the corresponding crosses. In addition, we have over-plotted known sources from the Magellanic Clouds classified by \cite[Woods et al. (2011)]{Woods_etal2011} and \cite[Ruffle et al. (2015)]{Ruffle_etal2015}, as well as post-AGB stars from our Galaxy \cite[(Szczerba et al. 2007;]{Szczerba_etal2007} \cite[2012)]{Szczerba_etal2012}. From the position of the HD evolutionary tracks and the over-plotted data, we are able to distinguish typical regions for AGB and post-AGB star candidates. 

Figure~\ref{fig:hess} presents the Hess diagram for 60,655 SMOG sources with AllWISE photometry with the A or B quality flag. 
The black lines are drawn according to the object distributions in Figure~\ref{fig:hd}. Below $[4.6]-[22.0]=6$ and to the right of the black lines we have AGB (both O- and C-rich) candidates (2,510 SMOG sources), while above $[4.6]-[22.0]=6$ we have post-AGB candidates (24,821 SMOG sources), since the theoretical tracks and most of the Galactic post-AGB candidates are located there. However, while the AGB candidates seem relatively easier to be isolated, the post-AGB candidates are mixed with the known YSOs. The additional data analysis and follow-up spectroscopic observations are necessary to confirm the evolutionary status of the post-AGB star (or PN) candidates. The final sample of AGB/post-AGB star candidates will be used to compare the stellar evolution in the Outer and Inner Galaxy.

\acknowledgements

Authors would like to acknowledge financial support from the National Science Center, Poland grants 2014/15/B/ST9/02111, 2011/01/B/ST9/02031 (R.Sz. and N.S.), and  2013/11/N/ST9/00400 (A.K.).


\begin{thebibliography}{}

\bibitem[Matsuura \etal (2014)]{Matsuura_etal2014}
{Matsuura, M., Bernard-Salas, J., Evans, T. L., et al.} 2014, 
{\textit MNRAS}, 439 1472

\bibitem[Ruffle \etal (2015)]{Ruffle_etal2015}
{Ruffle, P.M.E, Kemper, F., Jones,  O.C., et al.} 2015, 
{\textit MNRAS}, 451, 3504 

\bibitem[Szczerba \etal (2007)]{Szczerba_etal2007}
{Szczerba, R., Si\'odmiak, N., Stasi\'nska, et al.} 2007
\textit{Astronomy $\&$ Astrophysics}, 378, 465

\bibitem[Szczerba \etal (2012)]{Szczerba_etal_2012}
{Szczerba, R., Si\'odmiak, N., Stasi\'nska, G., et al.} 2012
\textit{Proceedings of the International Astronomical Union, IAU Symposium}, Volume 283, 506

\bibitem[Szczerba \etal (2016)]{Szczerba_etal2016}
{Szczerba, R., Si\'odmiak, N., Le\'sniewska, A., et al.} 2016,
 \textit{Journal of Physics: Conference Series}, Volume 728, article id. 042004
 
\bibitem[Steffen \etal (1998)]{Steffen_etal1998}
{Steffen, M., Szczerba, R., Schoenberner, D.}, 1998, 
\textit{Astronomy $\&$ Astrophysics}, 337, 149

\bibitem[Woods \etal (2011)]{Woods_etal2011}
{Woods, P. M., Oliveira,  J. M., Kemper, F., et al.} 2016, 
\textit{MNRAS}, 411, 1597
  



\end{thebibliography}
\end{document}